\documentclass[twocolumn,aps,prd,showpacs,superscriptaddress,floatfix,preprintnumbers]{revtex4}

\usepackage{graphicx}
\usepackage{amsmath,amssymb,latexsym}
\usepackage{bm}
\usepackage{epsfig}

\begin{document}

\title{Monopole condensation in two-flavour Adjoint QCD}

\author{Guido Cossu}
\affiliation{Scuola Normale Superiore and INFN, Pisa, Italy}
\author{Massimo D'Elia}
\affiliation{Dipartimento di Fisica and INFN, Genova, Italy}
\author{Adriano Di Giacomo}
\affiliation{Dipartimento di Fisica and INFN, Pisa, Italy}
\author{Giuseppe Lacagnina}
\affiliation{INFN sezione di Milano, Italy}
\author{Claudio Pica}
\affiliation{Brookhaven National Laboratory, Upton, NY 11973-5000, USA}

\preprint{IFUP-TH/2008-03, IFUM-912-FT}

\begin{abstract}
In QCD with adjoint fermions (aQCD) the deconfining transition takes place at a lower temperature than the chiral transition. We study the two transitions by use of the Polyakov Loop, the monopole order parameter and the chiral condensate.
The deconfining transition is first order, the chiral is a crossover. The order parameters for confinement are not affected by the chiral transition. We conclude that the degrees of freedom relevant to confinement are different from those describing chiral symmetry. 
\end{abstract}

\maketitle

\section{Introduction}

Deconfinement and chiral symmetry restoration are two important features of QCD. Despite the fact that these
two phenomena are in principle independent of each other, finite
temperature lattice simulations indicate that they
occur at the same temperature within errors \cite{Karsch:2007dp}, making it hard to disentangle
them.  In particular, it is not clear yet what is the interplay between
the degrees of freedom relevant for the two transitions.

Several proposals exist in the literature for the confinement dynamics,
most of which are based on the presence of topological excitations in
the theory. A possible mechanism for confinement is dual
superconductivity of the QCD vacuum, which manifests itself as
condensation of magnetic charges \cite{thooft}. In order to
investigate this property of the vacuum, one constructs an operator
which carries magnetic charge and determines its vacuum expectation
value, which is expected to be different from zero in the confined
phase and strictly zero in the deconfined symmetric phase
\cite{pisa1,pisa2, pisa3, pisa4,fullqcd1}. Contrary to what happens for the Polyakov Loop, the symmetry described by this order parameter is not spoiled by the presence of dynamical quarks. It can therefore be used as a good order parameter for confinement also in full QCD. It has been
shown indeed~\cite{pisa4,fullqcd1} that in QCD with fermions in the
fundamental representation dual superconductivity disappears at the
same temperature where the chiral-deconfinement phase transition takes place.

While there is much evidence that monopole condensation is strictly related to
the dynamics of colour confinement, a still unsolved issue concerns the
relation between dual superconductivity and the dynamics of chiral
symmetry breaking. As already pointed out, the coincidence of
deconfinement and chiral restoration makes this problem difficult in
ordinary QCD. A system where deconfining and chiral
transitions are distinct could provide the framework to investigate this
issue. One such system is QCD with quarks in the adjoint
representation of $SU(3)$ (aQCD), in which the two transitions
seemingly take place at different temperatures
\cite{karsch}. Furthermore, the coupling to the adjoint quarks does not
explicitly break the $Z(3)$ center symmetry of the action, and
therefore the two transitions can be characterized by two order
parameters, namely the Polyakov loop and the chiral condensate.

The authors of ref.~\cite{karsch} performed lattice simulations of
aQCD with two flavours of staggered quarks, and found two distinct
phase transitions, with $\beta_{\rm dec}<\beta_{\rm chiral}$. They
observed a strong first order deconfinement transition and a
continuous chiral transition. They also checked that the Polyakov Loop,
which is sensitive to deconfinement, is not significantly affected by
the chiral transition. The nature of the chiral transition in $N_f=2$
aQCD has been further investigated in \cite{engels}, where the authors
made an extensive analysis with the aim of determining the order of the
chiral transition. They found that the behaviour of the magnetic
equation of state was consistent with the presence of a second order
chiral transition in the zero quark mass limit (see also \cite{Basile:2004wa}).  For the purpose of this work, however, it is sufficient to know that a chiral transition exists and is separated from the deconfinement one.

The structure of the paper is the following. In Section \ref{aQCD_mono} we briefly review the basics of aQCD and of the order parameter for monopole condensation. In
Section \ref{simulations} we report the results of our study of the
deconfinement transition both by use of the magnetic order parameter
and of the Polyakov loop. We make a summary and draw conclusions in
Section \ref{conclusions}.

\section{Adjoint QCD and magnetic order parameter}
\label{aQCD_mono}

\subsection{aQCD}
Quarks in the adjoint representation of $SU(3)$ have $8$ color
degrees of freedom and can be described by $3\times 3$ hermitian
traceless matrices:
\begin{equation}
Q(x) = Q^a(x)\lambda_a
\end{equation}
where $\lambda_a$ are the Gell-Mann's matrices. In order to write the fermionic part of the action for this model, the $8-$dimensional $U_{(8)}$ representation
of the gauge links (which is real) must be used:
\begin{equation}
U^{ab}_{(8)} = \frac{1}{2}{\rm Tr}\left
(\lambda^aU_{(3)}\lambda^bU_{(3)}^{\dagger}\right )
\label{eq_u8}
\end{equation}
The full action is therefore given by:
\begin{equation}
S = \beta S_G[U_{(3)}] + \sum_{x,y} {\bar Q}(x)^a M^{ab}\left (U_{(8)}\right
)_{x,y}Q(y)^b
\end{equation}
where $S_G$ is the usual $SU(3)$ gauge action with links in the
3-dimensional representation and $M$ is the staggered fermions matrix. The Polyakov loop is defined as in the
pure gauge case:
\begin{equation}
L_{(3)} \equiv  \frac{1}{3L_s^3} \langle \left| \sum_{{\vec x}}{\rm
  Tr}\prod_{x_0=1}^{L_t}(U_{(3)})_0(x_0,{\vec x})\right| \rangle
\end{equation}
where $L_s$ and $L_t$ are the spatial and temporal sizes of the lattice respectively and the trace is over color indexes.
This quantity is an order parameter for the spontaneous breaking of
the center symmetry which is not broken by adjoint fermions (see
Eq. \ref{eq_u8}). A well known result \cite{svetitsky}
relates $L_{(3)}$ to the free energy of an isolated static quark in the fundamental representation
\begin{equation}
L_{(3)} \propto e^{-F/T}
\end{equation}
in a gluonic bath at temperature $T$. In the center symmetric phase,
where $L_{(3)}=0$, the free energy is infinite, thus realizing
confinement; $L_{(3)}$ can therefore be used as an order parameter for the
deconfinement transition.

\subsection{Monopole condensation}

The vacuum expectation value of a magnetically charged operator, $\langle \mu \rangle$, was proposed in references \cite{pisa1,pisa2, pisa3, pisa4} as an
order parameter for the deconfinement transition. The operator detects condensation of magnetic charges, i.e. Higgs breaking of the magnetic charge symmetry.
The procedure involves a gauge-fixing, the so-called Abelian Projection \cite{abelian}. However, the particular choice of the gauge is inessential as shown by numerical simulations \cite{pisa3} and by theoretical arguments \cite{theor}. The explicit form of the {\it vev} of the magnetic charged operator is given by
\begin{equation}
\langle \mu \rangle = \frac{1}{Z}\int [dU] e^{-{\widetilde
S}}=\frac{{\widetilde Z}}{Z}
\end{equation}
${\widetilde S}$ is obtained from the original action by the
insertion of a monopole field in the temporal plaquettes of a given
time slice \cite{pisa1}. The measurement of a ratio of partition functions
is a difficult numerical task and so, to better cope with fluctuations, one calculates the quantity
\begin{equation}
\rho = \frac{\partial}{\partial\beta} \ln \langle \mu \rangle =
\langle S_G\rangle_S - \langle\widetilde{S_G}\rangle_{\widetilde{S}}
\end{equation}
where $S_G$ is the ordinary gauge part of the action. Clearly, two simulations have to be run
for each value of $\beta$, with and without the monopole insertion. The
drop of the order parameter at the transition corresponds to a 
peak of $-\rho$. In the vicinity of the critical temperature a scaling
ansatz for the order parameter
\begin{equation}
\langle \mu \rangle \simeq L_s^{-\beta_\mu/\nu}f(L_s^{1/\nu}(\beta_c-\beta))
\end{equation}
($\beta_\mu$ is the critical exponent associated to the order
parameter) implies
\begin{equation}
\rho \simeq L_s^{1/\nu}f_\rho( L_s^{1/\nu}(\beta_c-\beta))
\end{equation}
where $\nu$ is the critical index of the correlation lenght and $f,f_\rho$ are universal scaling functions. In the case of a weak first order transition the critical exponent $\nu$ is equal to $1/3$, i.e. the $-\rho$ peak is expected to scale with the spatial volume:
\begin{equation}
\rho/L_s^3  \simeq f_\rho( L_s^{3}(\beta_c-\beta)).
\label{scaling_rho}
\end{equation}

\section{Simulations and results}
\label{simulations}

We simulated two flavours of adjoint staggered fermions using the
exact RHMC algorithm \cite{rhmc} for the simulations with the monopole
insertion and the $\Phi$ algorithm \cite{duane} for the other
simulations. Trajectories had a length of $N_{MD}\delta t=0.5$, and
typical integration steps $\delta t = 0.02, 0.005$ depending on the
mass (see below). Acceptance rate was above $80\%$ on average.  Inversions of the fermionic matrix were performed using the Conjugate Gradient
Algorithm. We have run simulations mostly with two different lattice sizes,
$L_s^3\times L_t=12^3\times 4, 16^3\times 4$, and bare quark masses
$am_q=0.01, 0.04$. We have evaluated the average plaquette, the $\rho$
parameter, the Polyakov loop, and the chiral condensate for several
values of $\beta$ in the range $(3.0, 8.0)$ (smallest $\beta$s are not shown in graphs). In order to simulate the action $\tilde S$, $C^*$ boundary conditions have been implemented \cite{cstar}. Our code has been run on the APEmille machine in Pisa and the apeNEXT facility in Rome.

\subsection{Results}

\begin{figure}
\includegraphics*[width=1.0\columnwidth]{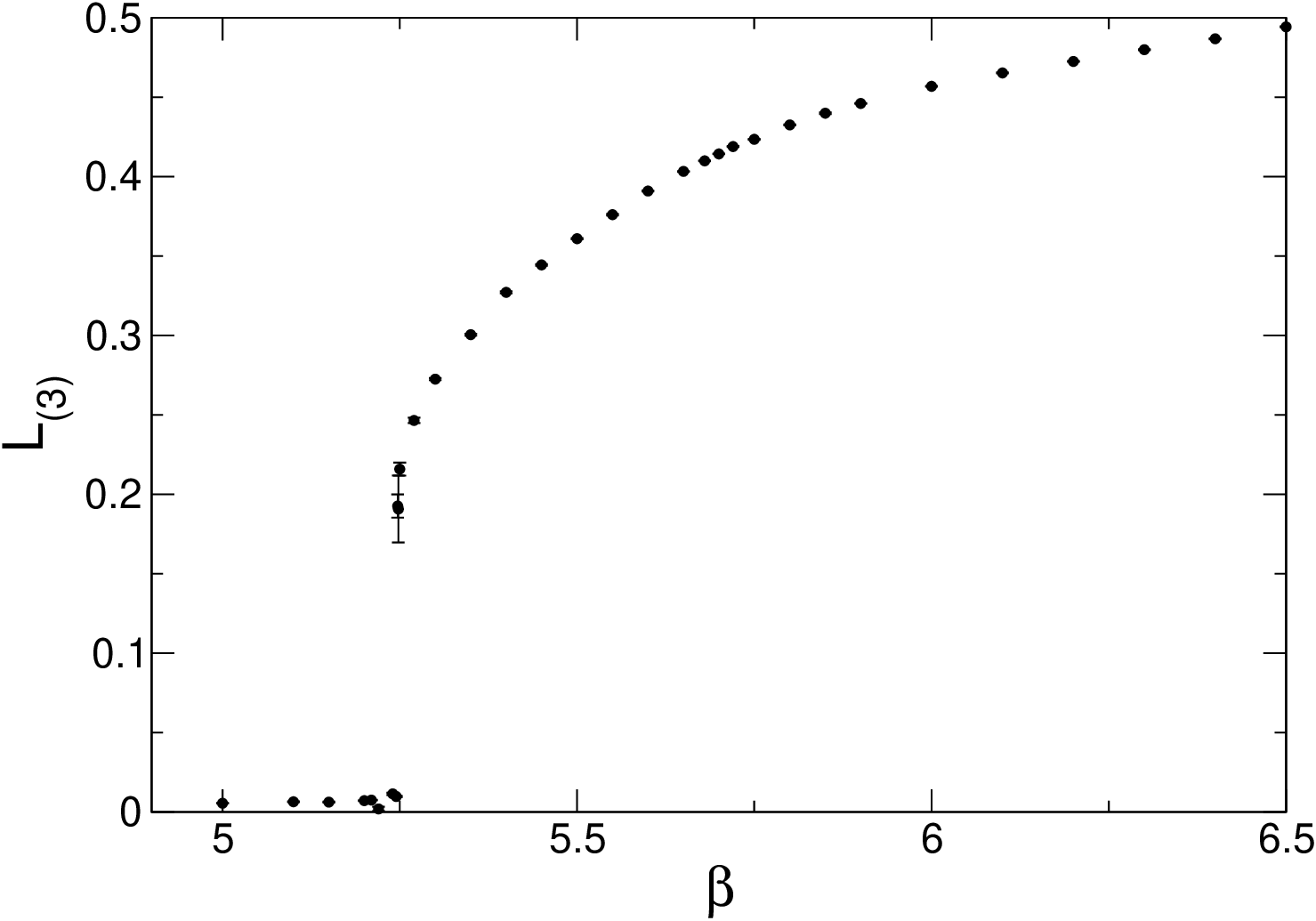}
\caption{The Polyakov loop, with $am=0.01$ and $16^3\times 4$
lattice.}
\label{polyakov}
\end{figure}

The thermodynamical properties of the Polyakov loop were the easiest
to study. We used this observable as a reference to investigate the confinement-deconfinement phase transition by means of the $\rho$ parameter. The Polyakov
loop shows the typical behaviour of a sharp first order transition
(see Fig. \ref{polyakov}). The pseudocritical value of $\beta$ for the
smallest mass was estimated, by inspection of
the data, to be at $\beta=5.25$ (independently of the volume) where $L_{(3)}$ shows a clear
discontinuity. This result is in agreement with \cite{karsch},
where a smaller volume $8^3\times 4$ and the same bare quark mass were
simulated. We shall use this pseudocritical $\beta$ as an estimate for
$\beta_{\rm dec}$ in our finite size scaling analysis.

\begin{figure}
\includegraphics*[width=1.0\columnwidth]{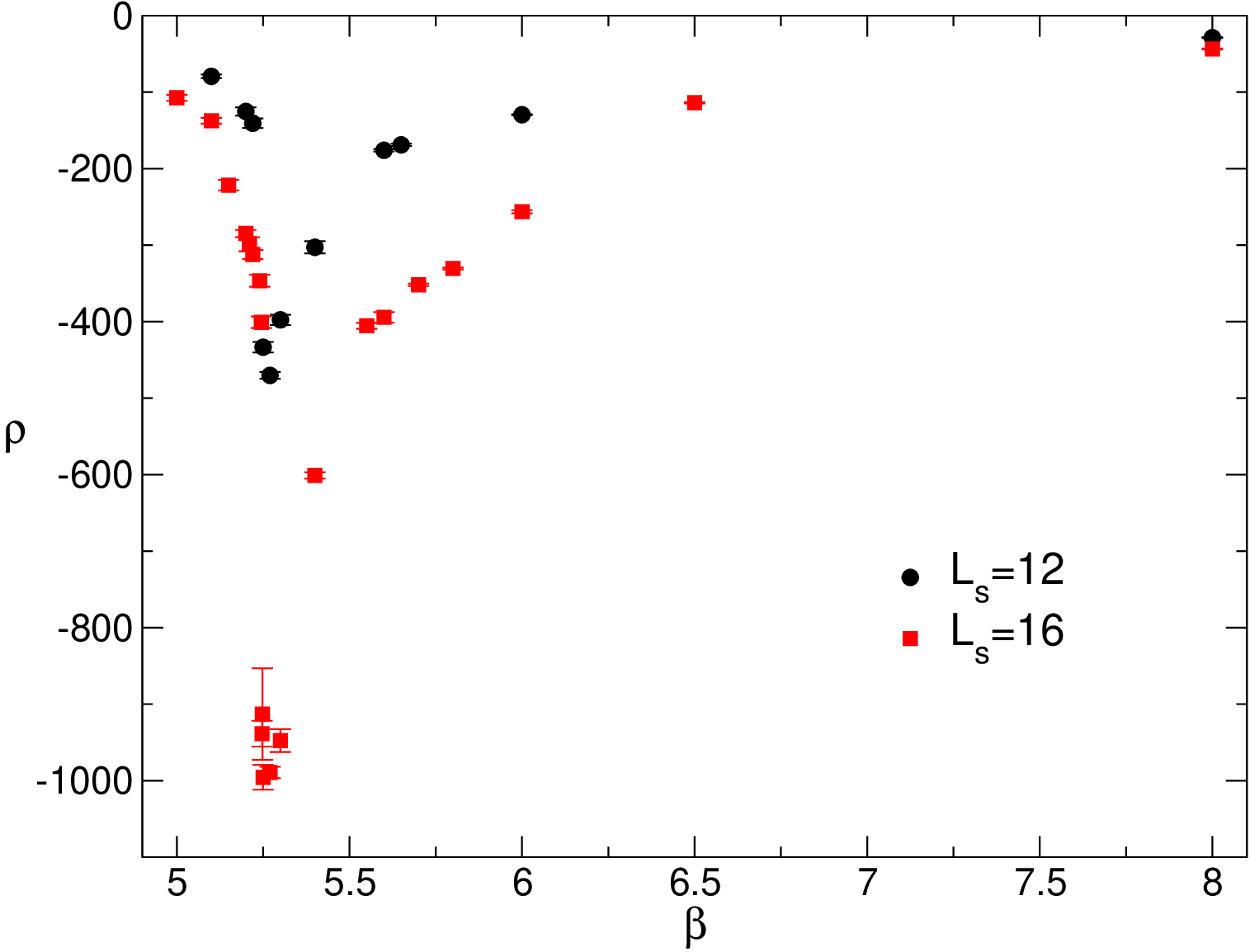}
\caption{The $\rho$ parameter, with $am=0.01$, $L_t=4$, for two different
  spatial volumes.}
\label{plot_rho}
\end{figure}

\begin{figure}
\includegraphics*[width=1.0\columnwidth]{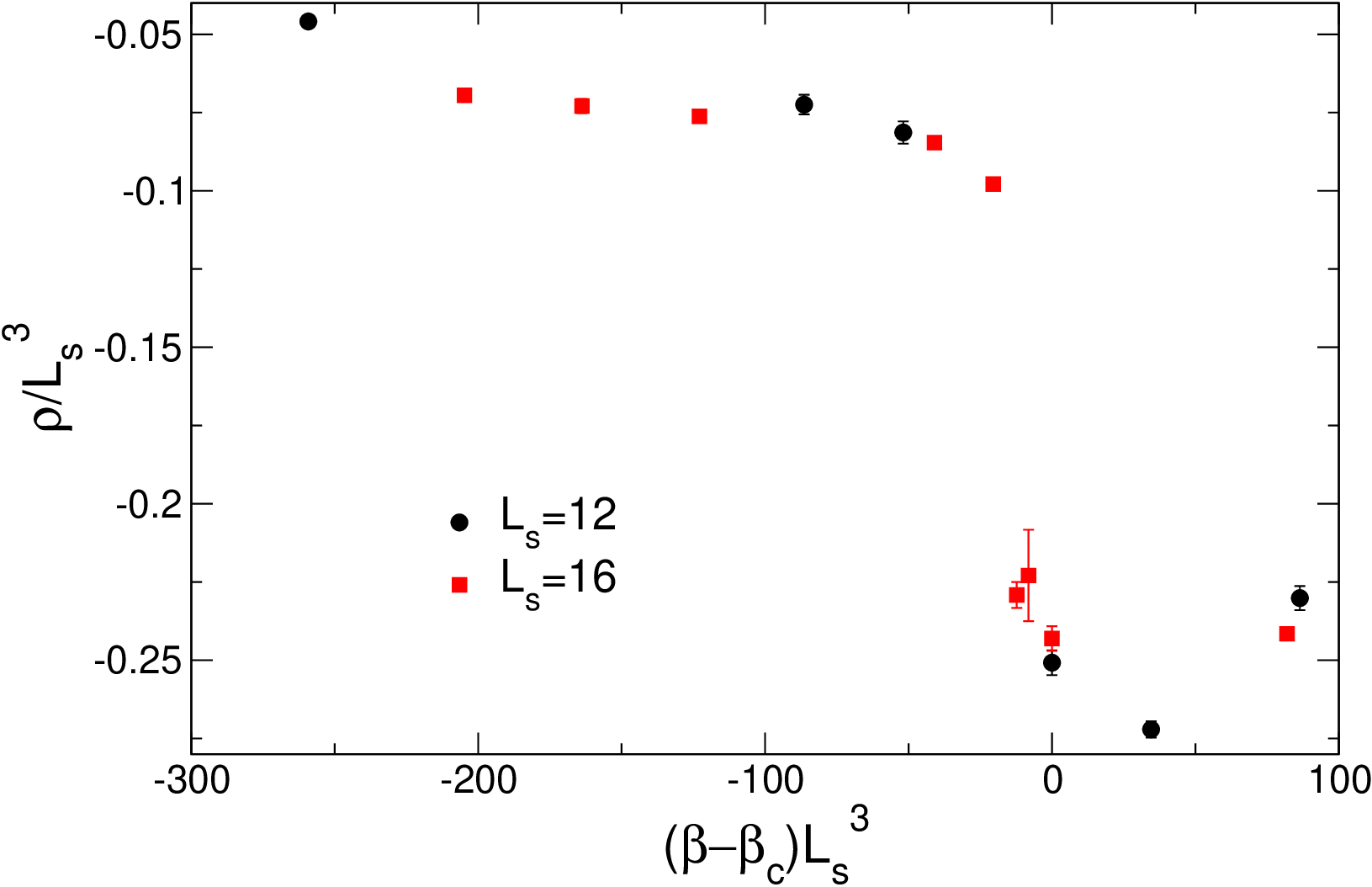}
\caption{Scaling of the $\rho$ parameter, $am=0.01$,
$L_t=4$. $\beta_c=5.25$, estimated from the Polyakov loop at $L_s=16$.}
\label{rho_scaling}
\end{figure}

\begin{figure}
\includegraphics*[width=1.0\columnwidth]{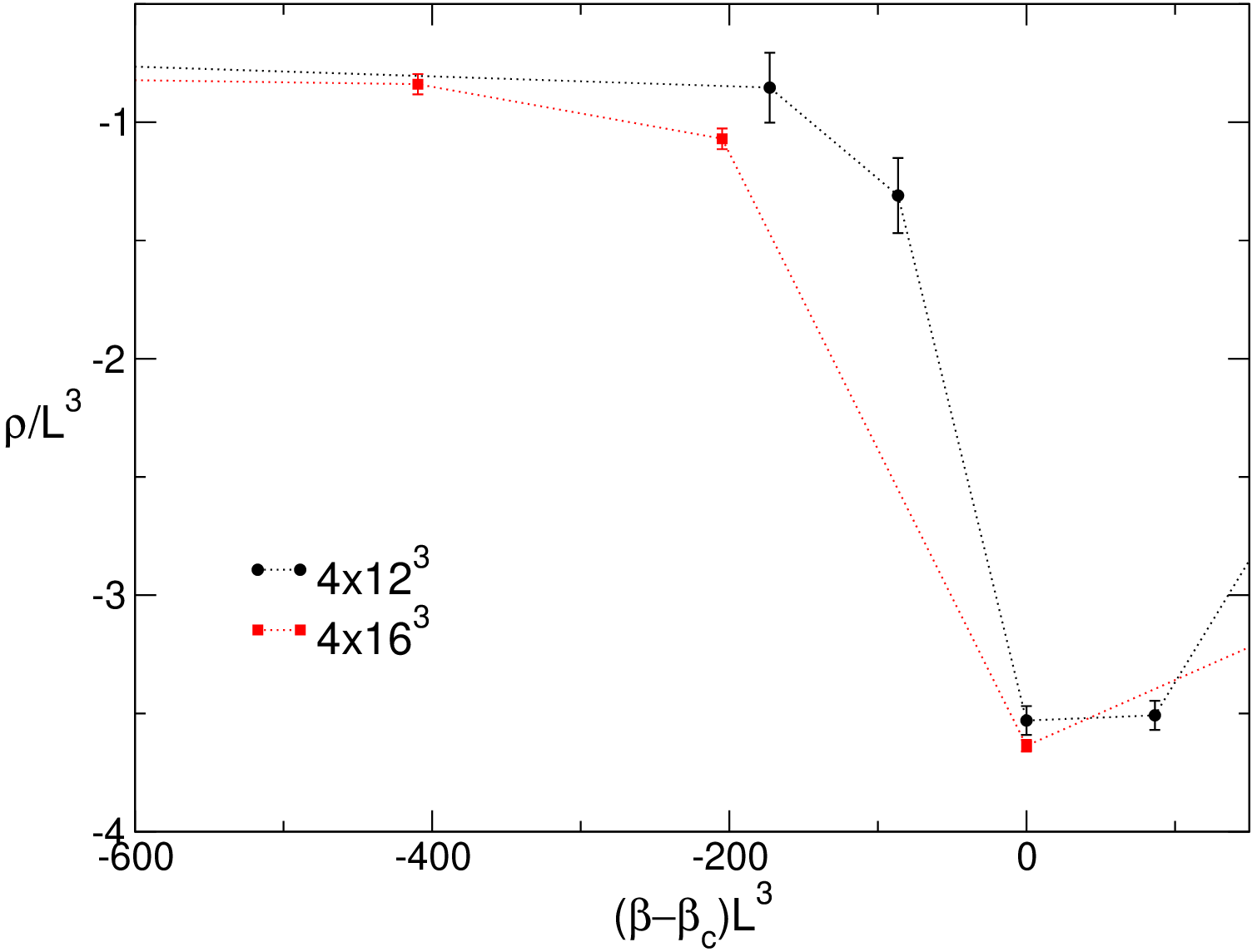}
\caption{Scaling of the $\rho$ parameter, $am=0.04$,
$L_t=4$. $\beta_c=5.25$, estimated from the Polyakov loop at $L_s=16$.}
\label{rho_scaling_004}
\end{figure}

\begin{figure}
\includegraphics*[width=1.0\columnwidth]{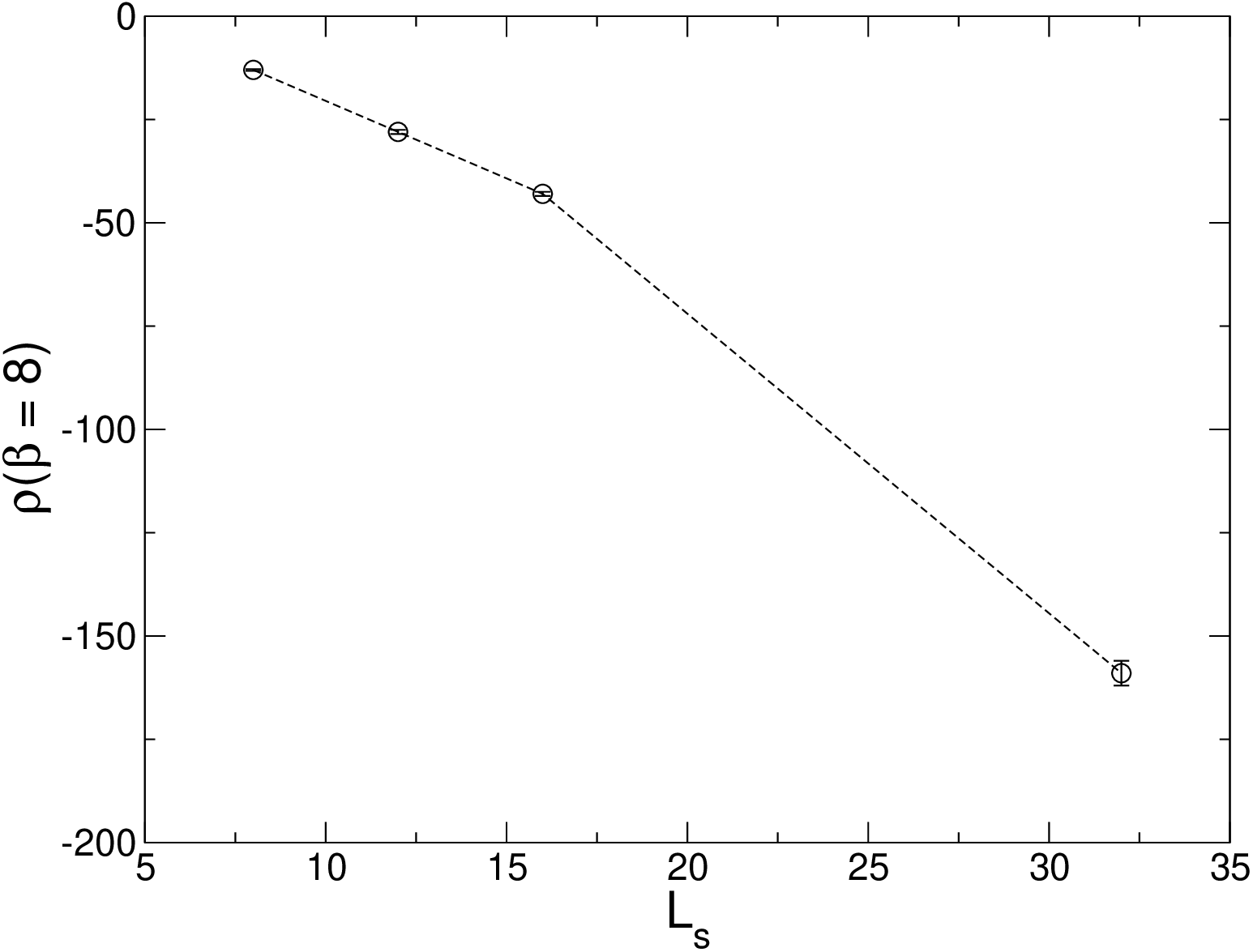}
\caption{Behavior of the $\rho$ observable varying the linear dimension of the lattice at fixed $\beta = 8$. }
\label{rho8}
\end{figure}

For the magnetic order parameter, we find the expected  peak
at values of $\beta$ which coincide within errors with those at which the
Polyakov loop has its discontinuity. The $\rho$ parameter is expected to approach zero
independently of the volume as $\beta \rightarrow 0$; it should also
diverge with $L_s$ in the deconfined region \cite{rho_prop}. We found a
good qualitative agreement of our data with the expectations in both limits (see Figs. \ref{plot_rho}, \ref{rho_scaling}, \ref{rho8}). Around the transition finite size
scaling analysis shows that $\rho$ has the scaling properties of a first order transition for both values of the
quark mass (Figs. \ref{rho_scaling}, \ref{rho_scaling_004}). In particular the height of the peak is proportional to $L_s^3$ within errors as expected from eq. \ref{scaling_rho}.

\begin{figure}
\includegraphics*[width=1.0\columnwidth]{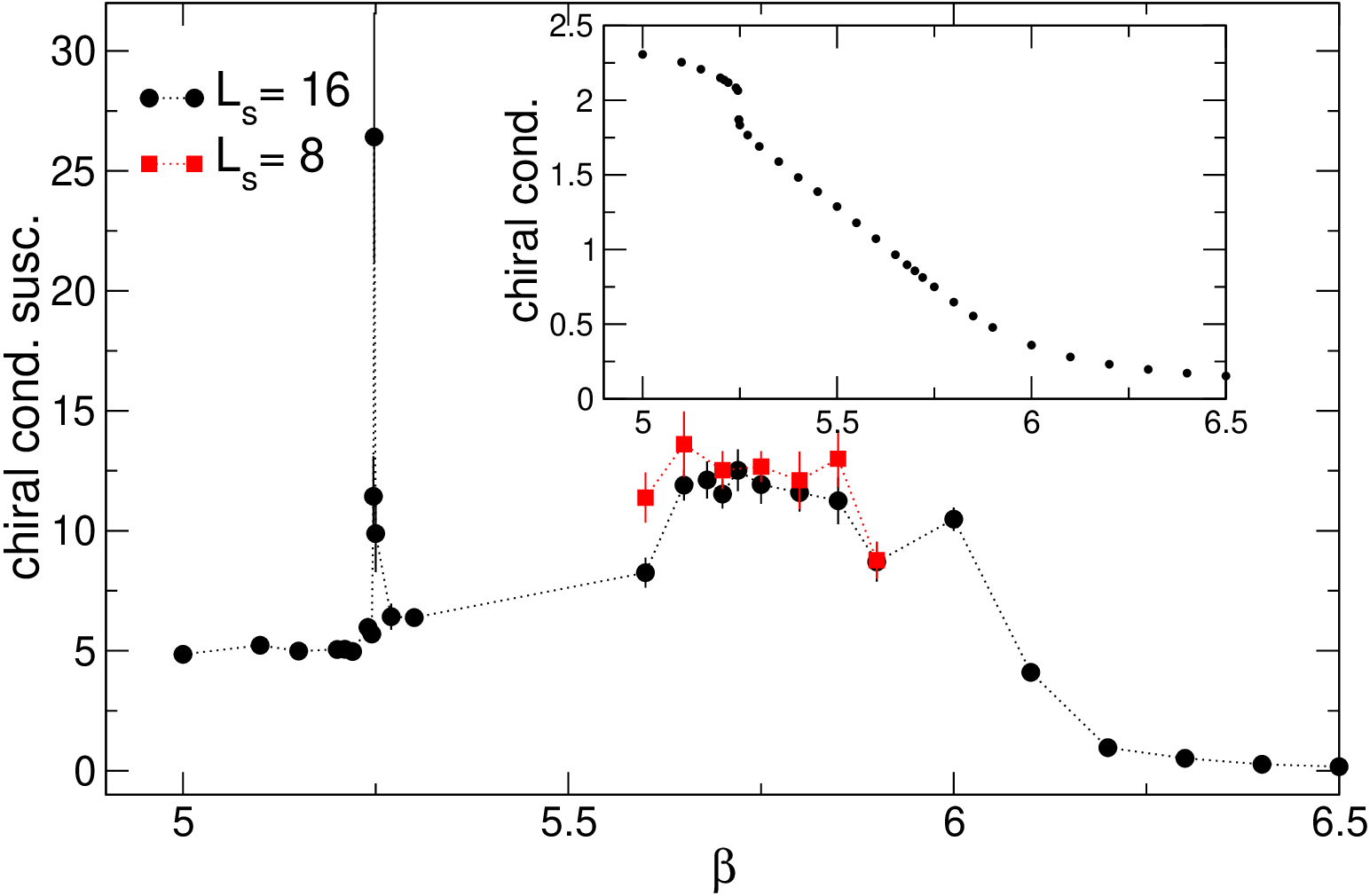}
\caption{Susceptibility of the chiral condensate, with $am=0.01$ and two different lattices: $8^3\times 4$, $16^3\times 4$. The data are compatible with a crossover at the chiral transition. In the inset, the chiral condensate within the same range of $\beta$ values. A clear jump is visible at the deconfinement phase transition.}
\label{chisusc}
\end{figure}

We also looked for possible effects of the chiral transiton on the magnetic order parameter. The first step to address
this issue was to locate the transition by means of the natural
order parameter, the chiral condensate $\langle {\bar
\psi}\psi\rangle$. This parameter and its susceptibility
(Fig. \ref{chisusc}) were found both consistent with a chiral
transition in the region around $\beta_{\rm chiral}=5.8$. An unambiguous peak in the susceptibility of the chiral condensate is visible only for the lightest mass, $am=0.01$. Comparison of results at different volumes shows that the chiral transition, at this value of the fermionic mass, is compatible with a crossover. These results are in agreement with those already contained in \cite{karsch, engels}. The $\rho$ parameter does not show any significant change at $\beta\simeq \beta_{\rm chiral}$
(Fig. \ref{plot_rho}), the same happens for the Polyakov Loop. Furthermore, the analysis of \cite{engels}
shows that a bare quark mass of $am=0.01$ is close enough to the
scaling region of the chiral transition. It is therefore safe to
conclude that the $\rho$ parameter is not affected by the chiral
transition, i.e. that different d.o.f. dominate at the two transitions.

\section{Summary and conclusions}
\label{conclusions}

We have studied deconfinement and the chiral transition in lattice QCD with two flavors in the adjoint representation. 

Deconfinement is detected as a sharp jump at the critical temperature $\beta_c$ of the Polyakov Loop (Fig. \ref{polyakov}). It is also seen as a sharp peak of the susceptibility $\rho$ related to magnetic charge condensation. 
The location of the jump and of the peak concide within errors. Both parameters obey the scaling of a first order phase transition. 

The chiral order parameter $\langle \bar \psi \psi \rangle$ has a drop at the deconfining transition, corresponding to a peak in its susceptibility (Fig. \ref{chisusc}) but does not vanish above it. $\langle \bar \psi \psi \rangle$ drops to zero at an higher temperature where its susceptibility $\chi$ has a broad peak  (chiral transition). The scaling of $\chi$ is compatible with a crossover. Neither the Polyakov line nor the magnetic order parameter show any change at the chiral transition. 

From these observations we can conclude that:

\begin{enumerate}
\item the magnetic order parameter detects deconfinement on the same footing as the Polyakov Loop. This again corroborates the mechanism of confinement by dual superconductivity of the vacuum, as in the case of pure gauge \cite{thooft,pisa1}.
\item The degrees of freedom relevant to deconfinement are different from those relevant to chiral transition. The detectors of deconfinement ($L_{(3)}$ and $\rho$) are insensitive to the chiral transition, and $\langle \bar \psi \psi \rangle$ is non zero above the deconfinement transition. 
\end{enumerate}

The last conclusion above can be relevant to the study of ordinary QCD with $N_f=2$ in the fundamental representation \cite{D'Elia:2005bv}. There the two transitions for some reason occour at the same $\beta_c$ and the interplay of the two different kinds of degrees of freedom could be at the origin of the difficulties in determining the order of the transition. The analysis based only on chiral degrees of freedom \cite{Pisarski:1983ms} might prove to be inadequate.

The work of C.P. has been supported by contract No. DE-AC02-98CH10886 with the U.S. Department of Energy. We wish to thank the apeNEXT staff in Rome for the support.


\begin{thebibliography}{99}

\bibitem{Karsch:2007dp}
  F.~Karsch, PoS(LATTICE 2007)015, arXiv:0711.0656; arXiv:0711.0661 [hep-lat] and references therein.
\bibitem{thooft} G.~'t Hooft, Nucl.\ Phys.\ B {\bf 190} (1981), 455
\bibitem{pisa1} A.~Di Giacomo, B.~Lucini, L.~Montesi, G.~Paffuti, Phys. Rev. D
  {\bf 61}, 034503 (2000), hep-lat/9906024
\bibitem{pisa2} A.~Di Giacomo, B.~Lucini, L.~Montesi, G.~Paffuti,
  Phys. Rev. D{\bf 61}, 034504 (2000), hep-lat/9906025
\bibitem{pisa3} J.~M.~Carmona, M.~ D'Elia, A.~Di Giacomo, B.~Lucini,
  G.~Paffuti, Phys. Rev. D{\bf}64, 114507 (2001), hep-lat/0103005
\bibitem{pisa4} M.~ D'Elia, A.~Di Giacomo, B.~Lucini, G.~Paffuti, C.~Pica,
  Phys. Rev. D{\bf 71}, 114502 (2005), hep-lat/0503035
\bibitem{fullqcd1} J.~M.~Carmona, M.~D'Elia, L.~Del Debbio, A.~Di
  Giacomo, B.~Lucini and G.~Paffuti, Phys. Rev. D {\bf 66}, 011503
  (2002), hep-lat/0205025
\bibitem{karsch} F.~Karsch, M.~Lutgemeier, Nucl. Phys. B {\bf 550}, 449
  (1999), hep-lat/9812023
\bibitem{engels} J.~Engels, S.~Holtmann, T.~Schulze, Nucl. Phys. B
{\bf 724}, 357 (2005), [arXiv:hep-lat/0505008].
\bibitem{svetitsky} L.~D.~McLerran, B.~Svetitsky, Phys. Rev. D
{\bf 24}, 450 (1981)
\bibitem{abelian} L.~Del Debbio, A.~Di Giacomo, B.~Lucini and
G.~Paffuti, hep-lat/0203023
\bibitem{theor} A.~Di Giacomo, G.~Paffuti, Nucl.\ Phys.\ Proc.\ Suppl. {\bf 129} (2004), 647
\bibitem{rhmc} M.~A.~Clark, A.~D.~Kennedy, Nucl. Phys. Proc. Suppl. {\bf 129},
  850 (2004), hep-lat/0309084
\bibitem{duane} S.~Duane, A.~D.~Kennedy, B.~J.~Pendleton and
  D.~Roweth, Phys.\ Lett.\ B {\bf 195}, 216 (1987)
\bibitem{cstar} J.~M.~Carmona, M.~ D'Elia, A.~Di Giacomo,
  B.~Lucini, Int. J. Mod. Phys. C{\bf 11}, 637 (2000), hep-lat/0003002
\bibitem{rho_prop} A.~Di Giacomo, G.~Paffuti, Phys.\ Rev.\ D {\bf 56}, (1997), 6816 
\bibitem{D'Elia:2005bv}
  M.~D'Elia, A.~Di Giacomo and C.~Pica,
  Phys.\ Rev.\  D {\bf 72}, 114510 (2005)
  [arXiv:hep-lat/0503030] and
  G.~Cossu, M.~D'Elia, A.~Di Giacomo and C.~Pica,
  arXiv:0706.4470 [hep-lat].
\bibitem{Pisarski:1983ms}
  R.~D.~Pisarski and F.~Wilczek,
  Phys.\ Rev.\  D {\bf 29}, 338 (1984).
\bibitem{Basile:2004wa}
  F.~Basile, A.~Pelissetto and E.~Vicari,
  JHEP {\bf 0502}, 044 (2005)
  [arXiv:hep-th/0412026].

\end{thebibliography}
\end{document}